# Nuclear structure studies performed using the ($^{18}$O,$^{16}$O) two-neutron transfer reactions

**D. Carbone[1], C. Agodi[1], F. Cappuzzello[1,3], M. Cavallaro[1], J. L. Ferreira[2], A. Foti[3,4], A. Gargano[5], S.M. Lenzi[6,7], R. Linares[2], J. Lubian[2], G. Santagati[1]**

[1]INFN Laboratori Nazionali del Sud, Catania, Italy
[2]Instituto de Fisica, Universidade Federal Fluminense, Niteroi, RJ, Brazil
[3]Dipartimento di Fisica e Astronomia, Università degli Studi di Catania, Italy
[4]INFN Sezione di Catania, Catania, Italy
[5]INFN Sezione di Napoli, Napoli, Italy
[6]Dipartimento di Fisica e Astronomia, Università di Padova, Padova, Italy
[7]INFN Sezione di Padova, Padova, Italy

carboned@lns.infn.it

**Abstract**. Excitation energy spectra and absolute cross section angular distributions were measured for the $^{13}$C($^{18}$O,$^{16}$O)$^{15}$C two-neutron transfer reaction at 84 MeV incident energy. This reaction selectively populates two-neutron configurations in the states of the residual nucleus. Exact finite-range coupled reaction channel calculations are used to analyse the data. Two approaches are discussed: the extreme cluster and the newly introduced microscopic cluster. The latter makes use of spectroscopic amplitudes in the centre of mass reference frame, derived from shell-model calculations using the Moshinsky transformation brackets. The results describe well the experimental cross section and highlight cluster configurations in the involved wave functions.

## 1. Introduction

Direct two-nucleon transfer reactions play an important role in the study of specific features of the atomic nucleus and indeed they were extensively explored [1] [2] [3] [4] [5] to study for example pairing correlations. Among these, heavy-ion direct transfer reactions at energies close to the Coulomb barrier are useful tools to obtain precise spectroscopic information.

In this context, a wide range of systems were explored by heavy-ion induced one- and two-neutron transfer reactions at INFN-LNS (Italy) by the ($^{18}$O,$^{17}$O) and ($^{18}$O,$^{16}$O) reactions. The MAGNEX spectrometer [6] was used to detect the ejectiles. The large acceptance and high resolution of it allowed to obtain high quality energy spectra up to the region above the two-neutron separation energy in the residual nucleus [7] [8] [9]. New phenomena were observed, e.g. the dominance of the direct one-step transfer of the two neutrons [10] and the presence of the first experimental signature of the Giant Pairing Vibration [11] [12] at high excitation energy in the $^{14}$C and $^{15}$C spectra. The analysis of the broad structures in the $^{15}$C spectrum at high excitation energy was presented in ref. [13] and the neutron decay of these structures was investigated in ref. [14], exploiting the use of the EDEN array coupled to the spectrometer [15] [16]. Moreover, it was demonstrated in several works that the ($^{18}$O,$^{16}$O) two-neutron transfer reaction can be used for quantitative spectroscopic studies of pair configurations in nuclear states [17] [18] [19].

A complete theoretical treatment of the transfer process should contain a description of three terms: the one-step channel with the inclusion of all the possible inelastic excitations of the target, projectile, ejectile or residual nucleus; the sequential channel, with the inclusion of intermediate partitions and the non-orthogonal term deriving from the limited model space used in actual calculations. However, in the state-of-the-art theories a complete model is still not available. As long as heavy-ions are concerned, it is necessary to include explicitly the inelastic excitations of the involved nuclei using the coupled-channels approach [20] [21] [22]. In the lack of a complete two-step Coupled Reaction Channel (CRC) theory, the one-step and two-step calculations can be performed separately and their amplitudes summed by considering the relative phase as an additional parameter.

In ref. [17] the experimental absolute cross sections of the one- and two- neutron transfer reactions induced by an $^{18}$O beam on a $^{12}$C target were reproduced without any scaling factor by means of Exact Finite Range (EFR) CRC calculations. Two approaches were used: the extreme cluster model and the independent coordinate scheme. The description of these approaches can be found in refs. [23] and [24], respectively. The same framework was recently applied to describe the absolute cross section angular distributions of the states populated by the $^{13}$C($^{18}$O,$^{16}$O) reaction [19]. A new approach consisting in a fully microscopic cluster calculation is introduced, performed by using the two-neutron spectroscopic amplitudes calculated in the shell-model framework. This new approach can be more extensively used with respect to the extreme cluster model to evaluate the presence of cluster components in the involved wave functions.

## 2. Experimental results

The experiment was performed at the INFN-LNS laboratory using an $^{18}$O$^{6+}$ beam at 84 MeV incident energy on a thin $^{13}$C target. The same measurement was also done on $^{12}$C target to estimate the background coming from the $^{12}$C impurities in the $^{13}$C target. The MAGNEX spectrometer was used to detect the ejectiles [6]. The particle identification and the data reduction technique are the same described in detail in refs. [12] [25]. An example of the obtained energy spectra for the $^{15}$C nucleus is shown in Fig. 1 (left panel), in which the $^{14}$C background spectrum coming from the $^{12}$C impurities in the $^{13}$C target is superimposed, after normalization. The strongest bound and resonant states appearing in the spectrum of the $^{15}$C nucleus are almost the same strongly populated in the (t,p) reactions on $^{13}$C [26]. In particular, below the one-neutron separation energy ($S_n$ = 1.218 MeV) the only two $^{15}$C bound states are identified, i.e. the ground ($J^\pi$ = 1/2$^+$) and the 5/2$^+$ state at $E_x$ = 0.74 MeV, that are characterized by a dominant single particle configuration [27]. In the region between $S_n$ and the two-neutron separation energy ($S_{2n}$ = 9.39 MeV), narrow resonances at $E_x$ = 3.103 (1/2$^-$), 4.22 (5/2$^-$), 4.66 (3/2$^-$), 6.84 (9/2$^-$, 7/2$^-$), 7.35 (9/2$^-$, 7/2$^-$) MeV are populated. All of these states are indicated to consist mainly of 2$p$-1$h$ configurations (with respect to the $^{14}$C$_{g.s.}$ vacuum state) [26]. Resonances with a single-particle configuration of a $^{14}$C$_{g.s.}$ + 1$n$ [28] are very weakly populated in the present reaction.

Examples of the obtained absolute cross section angular distributions for the ground state and the resonance at 4.22 MeV are shown in Fig. 1 (right panel).

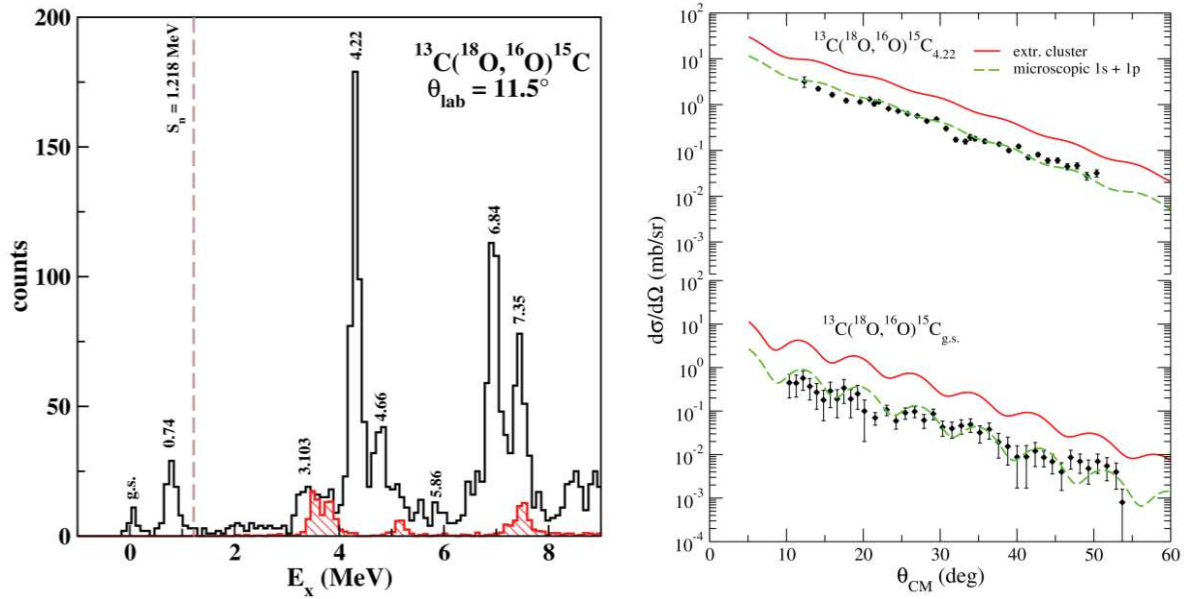

Fig. 1. (left panel) Excitation energy spectrum of the $^{13}$C($^{18}$O,$^{16}$O)$^{15}$C reaction for $10° < \theta_{lab} < 11°$. The red-hatched area corresponds to the background that comes from $^{12}$C impurities in the target. (right panel) Experimental cross section angular distributions for the g.s. and the state at 4.22 MeV in $^{15}$C. Theoretical calculations: extreme cluster calculations (red line) and microscopic cluster model calculations including 1s + 1p waves (green dashed line).

## 3. Theoretical analysis

Exact finite range coupled reaction channel (CRC) and two-step distorted wave Born approximation (DWBA) calculations for two-neutron transfer reactions were performed to describe the cross section, using the FRESCO code [29], with the same ingredients of ref. [17]. A direct, simultaneous transfer of the two particles was considered. The two-particle wave functions for the one-step two-neutron transfer mechanism were obtained considering two different schemes: i) the extreme cluster model, in which the relative motion between the two transferred neutrons is frozen and separated from the core and the two neutrons are coupled antiparallel to an intrinsic angular momentum $S = 0$, with spectroscopic amplitudes for both target and projectile set to 1.0 and ii) the *microscopic cluster model*, which will be introduced in Section 3.1. The $^{12}$C is assumed as a closed core and the model space includes the $1p_{1/2}$, $1d_{5/2}$, $2s_{1/2}$ orbits. The effective phenomenological *zbm* interaction [30] was used.

When a cluster model is used, the parameters relevant for the definition of the wave function are the principal quantum number $N$ and the orbital angular momentum $L$ relative to the core. These parameters are obtained from the conservation of the total number of quanta in the transformation of the wave function of two independent neutrons in orbits $(n_i, l_i)$ ($i = 1, 2$) into a cluster with internal state $(n, l)$ [31]: $2(n_1 - 1) + l_1 + 2(n_2 - 1) + l_2 = 2(N - 1) + L + 2(n - 1) + l$. In the extreme cluster model hypothesis, we consider $n = 1$ and $l = 0$ so that the cluster is in a 1s internal state. The spectroscopic amplitudes for both target and projectile were set to 1.0. The resulting differential cross sections using the extreme cluster hypothesis for the ground and the state at 4.22 MeV are shown in Fig. 1 (right panel). Using this extreme approximation the calculations gives a cross section larger than the experimental one. The main reason for this overestimation might lie in the approximation that the two neutrons are coupled to the total spin $S = 0$ with 100% of probability.

*3.1. The microscopic cluster model*

A natural way to go beyond the assumptions of the extreme cluster model is to introduce both parallel and antiparallel couplings for the two-neutrons. Realistic spectroscopic amplitudes are required for all the possible combinations of single-particle configurations in this enlarged space that can be derived from shell-model calculations. To achieve this goal, we made use of transformations from individual (*j-j* coupling) to relative and centre of mass coordinates (*LS* coupling) for the harmonic-oscillator wave functions of the two-particle system [19], using the Moshinsky brackets [31]. This approach is what we call *microscopic cluster calculations*.

As a first step, we performed microscopic cluster calculations considering that the cluster relative motion state is represented exclusively by $n = 1$ and $l = 0$ quantum numbers, i.e. the cluster is in the $1s$ intrinsic state. The obtained cross sections are much lower than data for all the states. Thus, we included also the $1p$ ($n = 1$ and $l = 1$) cluster relative motion states and the results are shown in Fig. 1 (right panel). We see that the results of $1s + 1p$ microscopic cluster calculations are in rather good agreement with the experimental angular distributions for both the ground and the state at 4.22 MeV.

To conclude, the new microscopic cluster model allows to describe rather well the experimental cross section for the states populated by the $^{13}C(^{18}O,^{16}O)^{15}C$ reaction, thus demonstrating the importance of a two-neutron correlation in the nuclear wave function in the two-neutron transfer mechanism. A dominance of the $1s$ and $1p$ waves in the two-neutron cluster internal wave functions is found.

**Acknowledgments**
This project has received funding from the European Research Council (ERC) under the European Union's Horizon 2020 research and innovation programme (grant agreement No 714625).